%% file: paper.tex
\documentclass[aps,pre,twocolumn,amsmath,amssymb,superscriptaddress,showpacs,longbibliography]{revtex4-1}

\usepackage{graphicx}
\usepackage{dcolumn}
\usepackage{bm}
\usepackage{mathrsfs}

\usepackage{natbib}
\usepackage[text={7in,9.5in},centering]{geometry}

\usepackage{hyperref}
\usepackage{url}

\usepackage{epsfig}
\usepackage{amsmath}
\usepackage{amssymb}
\usepackage{textcomp}

\usepackage{color}
\usepackage{float}

\begin{document}

\title{Choosing among alternative histories of a tree}

\author{G. Tim\'ar}
 \affiliation{Departamento de F\'\i sica da Universidade de Aveiro \& I3N, Campus Universit\'ario de Santiago, 3810-193 Aveiro, Portugal}

\author{R. A. da Costa}
 \affiliation{Departamento de F\'\i sica da Universidade de Aveiro \& I3N, Campus Universit\'ario de Santiago, 3810-193 Aveiro, Portugal}

\author{S. N. Dorogovtsev}
 \affiliation{Departamento de F\'\i sica da Universidade de Aveiro \& I3N, Campus Universit\'ario de Santiago, 3810-193 Aveiro, Portugal}

\author{J. F. F. Mendes}
 \affiliation{Departamento de F\'\i sica da Universidade de Aveiro \& I3N, Campus Universit\'ario de Santiago, 3810-193 Aveiro, Portugal}

\date{\today}

\begin{abstract}

The structure of an evolving network contains information about its past. Extracting this information efficiently, however,
is, in general, a difficult challenge. We formulate a fast and efficient method to estimate the most likely history of growing
trees, based on exact results on root finding. We show that our linear-time algorithm produces the exact stepwise most probable history in
a broad class of tree growth models. Our formulation is able to treat very large trees and therefore allows us to make reliable numerical
observations regarding the possibility of root inference and history reconstruction in growing trees. We obtain the general
formula $\langle \ln \mathcal{N} \rangle \cong N \ln N - cN$ for the size-dependence of the mean logarithmic number of possible histories of a
given tree, a quantity that largely determines the
reconstructability of tree histories. We also reveal an uncertainty principle: a relationship between the inferrability of the root and that
of the complete history, indicating that there is a tradeoff between the two tasks; the root and the complete history cannot both be inferred
with high accuracy at the same time.

\end{abstract}

\maketitle

\section{Introduction}
\label{sec1}

Models of network evolution have been studied intensively in the past two decades, and have resulted in an increasingly thorough understanding of the structure and function of various classes of real-world networks, both natural and artificial \cite{dorogovtsev2002evolution,newman2018networks}. Most theoretical works on evolution mechanisms have followed a deductive approach, focusing on structures resulting from general network evolution rules. The preferential attachment mechanism, made famous by Barab\'asi and Albert \cite{barabasi1999emergence}, has emerged as the top candidate to explain the structure of a multitude of observed networks with approximately power-law degree distributions. Well-known examples include the Internet \cite{faloutsos1999power}, the World Wide Web \cite{barabasi2000scale}, online social networks \cite{mislove2007measurement}, citation networks \cite{price1965networks} and networks of protein interactions \cite{jeong2001lethality}.
Various generalizations of the original preferential attachment model have been proposed (see for example \cite{krapivsky2000connectivity,bianconi2001competition,dorogovtsev2000evolution,dorogovtsev2000structure}), most notable of which uses a preference function that is an arbitrary linear function of node degree \cite{dorogovtsev2000structure}. This generalization of the preferential attachment model generates scale-free networks with degree distribution exponent in the range $\gamma \in ]2, \infty]$, thereby covering the vast majority of empirically observed exponents.

With the increasing availability of large-scale empirical network datasets it is becoming progressively easier to identify underlying network evolution mechanisms. Due to the abundance of large datasets the last decade has seen a surge of scientific activity on statistical inference methods. This has led to the emergence of the field of \emph{network archeology} \cite{navlakha2011network}, where the aim is to infer information about the history of a network from a current static snapshot of its structure. Such information can help us better understand its current structure and predict future states.

Inferring the history of growing networks, even considering the simplest growth mechanisms and network structures, is a difficult combinatorial problem. In recent years considerable literature has accumulated around the problem of inferring the first node in a growing network. Shah and Zaman \cite{shah2011rumors} formulated this problem in the language of finding the source of a spreading process on an underlying network, i.e., finding the first node of a growing infection network on an underlying network of possible transmissions. They proposed a novel measure of node centrality, which they call \emph{rumor centrality}, that counts the number of possible histories of the spreading process, assuming that the given node was the source. They showed that the rumor center (the node with the highest rumor centrality) is the maximum likelihood estimate of the source in the case where the underlying network is a Bethe lattice, i.e., an infinite $k$-regular network. Further, they showed that the probability of correct source detection in this case is non-zero in the infinite size limit. In follow-up work \cite{shah2016finding} they extend their results of non-zero source detection probability of the rumor center to heterogeneous random expanding trees, geometric trees, and heterogeneous spreading times. Similar methods have been used to identify multiple spreading sources \cite{luo2013identifying} and to detect the source given a set of viable candidates \cite{dong2013rooting}.

Another line of work deals directly with the problem of finding the root of growing trees. Bubeck et al. \cite{bubeck2017finding} studied root finding in random recursive trees and trees grown according to proportional preferential attachment. They showed that, for any desired probability, even as the tree size approaches infinity, a finite number of nodes can be identified that contain the root with that given probability. The authors studied the sizes of such confidence sets for root finding algorithms that rank nodes according to their rumor centrality or according to the size of the largest branch emanating from them. This ``maximum branch'' algorithm was also studied recently in the context of finding the seed - a small initial subtree - of random recursive trees in \cite{lugosi2019finding,reddad2019discovery}.

A far more complex problem is the inference of the complete history of a growing network. This problem was first considered explicitly in \cite{navlakha2011network}, where a greedy algorithm is introduced to reconstruct, in reverse, the evolution of protein-protein interaction networks, based on an approximation of the maximum a posteriori estimate of previous states.
Very recently more principled reconstruction methods have been suggested, based on a Bayesian inference framework. Magner et al. \cite{magner2018times,sreedharan2019inferring} estimate partial orders of node arrivals in a network, and give an upper bound on the attainable accuracy of partial order estimates. Such an upper bound on the accuracy is the result of inherent symmetries in graphs: there are generally many groups of structurally equivalent nodes whose arrival order cannot be determined. A different method attempts to estimate the expected arrival time of individual nodes in a network averaged over all possible histories \cite{young2019phase}.
Both of the above approaches rely on Monte Carlo sampling from an appropriately weighted distribution of possible histories, which are computationally demanding and hence not easily scalable. Simpler heuristic approaches were also suggested in \cite{young2019phase,sreedharan2019inferring}, based on node degrees and other simple centrality measures that are efficiently retrieved from network structure. Even more recently Cantwell et al. presented a message-passing method to calculate exactly the distribution of possible arrival times of nodes in growing trees of a broad class of growth models \cite{cantwell2019recovering}.

Despite considerable advances in the last decade, network archeology is still in its infancy and many avenues of investigation remain unexplored. The main purpose of this paper is to investigate the possibility of history reconstruction on growing trees using efficient, scalable algorithms. We discuss various existing methods, and also formulate a simple and fast tree history reconstruction algorithm exploiting the idea put forward by the seminal paper \cite{shah2011rumors}: counting the number of possible link sequences that could have generated a given tree. Our method gives a stepwise maximum likelihood estimate for the history in growth models where all possible histories of a tree have the same probability of occurring. Such models include linear preferential attachment, for example, which results in scale-free trees with degree distribution exponent in the range $]2, \infty]$. Despite the combinatorially large quantities involved, our approach is able to conveniently deal with very large networks. Our formulation also sheds light on the connection of the inference problem to the nonbacktracking matrix of the given tree. We give an informative comparison of different reconstruction methods for a wide range of recursive tree models. Our method enables us to numerically determine the mean logarithmic number of possible histories of large trees. We obtain the general functional form $\langle \ln \mathcal{N} \rangle \cong N \ln N - cN$ in terms of system size $N$, which is valid in a broad range of growth models. We also find an interesting relationship between the history degeneracy and the inferrability of the root in a wide range of growing trees, indicating that the easier it is to infer the root, the harder it is to reconstruct the complete history.

The outline of the paper is as follows. In Section \ref{sec2} we introduce useful concepts and discuss some basic symmetry properties of growing trees in general. In Section \ref{sec3} we formulate the root finding problem as a set of message passing equations that directly lead to a simple expression using the nonbacktracking matrix of a given tree. In Section \ref{sec4} we show that this approach is exact for general linear preferential attachment trees, where the preference function is an arbitrary linear function of node degree. We present numerical results for the efficiency of root finding in scale-free trees in Section \ref{sec5}. Based on the root finding scheme, in Section \ref{sec6} we propose a fast algorithm to reconstruct the complete history of a growing tree. We show that our method is exactly a stepwise maximum likelihood estimate of the history for linear preferential attachment trees. In Section \ref{sec7} we study the efficiency of history reconstruction, and accurately determine the functional form of the mean logarithmic history degeneracy in a wide range of growing trees. We also find an interesting scaling relationship between history degeneracy and the inferrability of the root, which is approximately valid in the entire range of recursive tree models considered. We give our conclusions in Section \ref{sec8}.

\section{Basic symmetry properties of growing trees}
\label{sec2}

In this section we introduce some definitions and discuss some properties of growing trees that will be useful in subsequent sections. For the most part we will be concerned with labeled trees, which we denote by $\mathcal{T}$. We assume that the nodes are arbitrarily labeled by numbers $1, 2, \ldots, N$, $N$ being the total number of nodes in the tree.
Let $\mathcal{S}(\mathcal{T})$ be the set of all labeled trees that are isomorphic to $\mathcal{T}$, in other words, the set of all labeled trees that share the structure of $\mathcal{T}$.
We define $\textrm{AUT}(\mathcal{T})$ to be the automorphism group of $\mathcal{T}$, i.e., the set of permutations of node labels that are automorphisms of tree $\mathcal{T}$. Then

\begin{equation}
|\mathcal{S}(\mathcal{T})| = \frac{ N! }{ |\textrm{AUT}(\mathcal{T})| },
\label{eq:2.1}
\end{equation}

\noindent
i.e., the number of distinct labeled trees that have the same structure $\mathcal{S}(\mathcal{T})$ is given by the number of possible permutations of the node labels ($N!$), divided by the number of label permutations that result in identical labeled trees ($|\textrm{AUT}(\mathcal{T})|$).

Let us now discuss some properties that are specific to growing trees. The complete evolution of a recursive tree built from $N$ labeled nodes can be imagined as a sequence of directed links $(i \leftarrow j)$ indicating that this particular link connects the new node $j$ to the already existing node $i$. The history of a recursive tree is thus completely described by a sequence of $N-1$ directed links, with the only criterion that each link must connect a new node to an existing one, that is, for each new link $(i \leftarrow j)$ node $i$ must already exist in the tree and node $j$ must not. Clearly, any given tree $\mathcal{T}$ can be generated by various different sequences, and to extract information about these possible sequences is exactly the task of history reconstruction. Let us denote the number of distinct link sequences that result in the same tree $\mathcal{T}$ by $\mathcal{N}_{\mathcal{T}}$. Also, let $\mathcal{N}_{\mathcal{S}(\mathcal{T})}$ denote the number of distinct link sequences that result in any labeled tree which is isomorphic to $\mathcal{T}$ (i.e., has the structure of $\mathcal{T}$). It is easy to see, from Eq. (\ref{eq:2.1}), that

\begin{equation}
\mathcal{N}_{\mathcal{S}(\mathcal{T})} = \mathcal{N}_{\mathcal{T}} \frac{ N! }{ |\textrm{AUT}(\mathcal{T})| }.
\label{eq:2.2}
\end{equation}

It is also informative to write the total number $\mathcal{N}(N)$ of distinct link sequences that result in any labeled tree of $N$ nodes. The root of the tree (the end-node of the first link $(i \leftarrow j)$) can be any of the $N$ possibilities. The second node (start-node of the first link) can be any of the remaining $N-1$ nodes. The end-node of the second link must be one of the existing $2$ nodes, and the start-node of the second link can be any of the remaining $N-2$. The $t^{\textrm{th}}$ link in the sequence must have, as its end node, one of the existing $t$ nodes, and the start node may be chosen from the remaining $N-t$. Therefore

\begin{equation}
\mathcal{N}(N) = N \prod_{t=1}^{N-1} (N-t)t = N \left[ (N-1)! \right] ^2.
\label{eq:2.3}
\end{equation}

\noindent
These sequences can have different probabilities of occurring in different growth models. In fact, we can define - as is often done in graph theory - a growth model as a distribution (a probability mass function) $\mathcal{M}(\vec{s})$ over all possible sequences $\vec{s} \in \{ (i_1 \leftarrow j_1), (i_2 \leftarrow j_2), \ldots, (i_{N-1} \leftarrow j_{N-1}) \}$. The simplest of such growth models is that of \emph{random recursive} (RR) trees, in which each new node attaches to a uniformly randomly chosen existing node. The distribution $\mathcal{M}(\vec{s})$ in this case is, clearly, uniform.
We can write the occurrance probability of any tree $\mathcal{T}$ of $N$ nodes in the RR model,

\begin{equation}
p_{\textrm{RR}, \mathcal{T}} = \frac{\mathcal{N}_{\mathcal{T}}}{ N \left[ (N-1)! \right] ^2 },
\label{eq:2.4}
\end{equation}

\noindent
and the occurrance probability of the structure of a given tree $\mathcal{T}$, as

\begin{equation}
p_{\textrm{RR}, \mathcal{S}(\mathcal{T})} = \frac{\mathcal{N}_{\mathcal{S}(\mathcal{T})}}{ N \left[ (N-1)! \right] ^2 } = \frac{\mathcal{N}_{\mathcal{T}}}{ (N-1)! \, |\textrm{AUT}(\mathcal{T})| },
\label{eq:2.5}
\end{equation}

\noindent
where we used Eq. (\ref{eq:2.2}).

An important class of growth models has the property that any sequence resulting in a given tree $\mathcal{T}$ has the same occurrance probability. We will refer to this as the \emph{equiprobable sequences} (EPS) property. Clearly, RR trees have the EPS property. In Section \ref{sec4} we will show this property also in the case of a wider class of growing trees. In a model that has the EPS property, we denote by $\pi_{\mathcal{T}}$ the occurrance probability of each sequence that results in labeled tree $\mathcal{T}$. Assuming that the growth model is independent of the (arbitrarily chosen) node labels, $\pi_{\mathcal{T}}$ is also the occurrance probability of any sequence resulting in the structure of $\mathcal{T}$ (i.e., any labeled tree isomorphic to $\mathcal{T}$.) We can write the occurrance probability of a given tree $\mathcal{T}$ in such models as

\begin{equation}
p_{\textrm{EPS},\mathcal{T}} = \pi_{\mathcal{T}} \mathcal{N}_{\mathcal{T}},
\label{eq:2.6}
\end{equation}

\noindent
and the occurrance probability of the structure of a given tree $\mathcal{T}$, as

\begin{equation}
p_{\textrm{EPS},\mathcal{S}(\mathcal{T})} = \pi_{\mathcal{T}} \mathcal{N}_{\mathcal{S}(\mathcal{T})} = \pi_{\mathcal{T}} \mathcal{N}_{\mathcal{T}} \frac{ N! }{ |\textrm{AUT}(\mathcal{T})| }.
\label{eq:2.7}
\end{equation}

\section{Root inference}
\label{sec3}

Given an arbitrary tree $\mathcal{T}$ and a growth model $\mathcal{M}$, we derive an expression for the probabilities $P_{\mathcal{M}}(i | \mathcal{T})$ that node $i$ was the root of the tree, assuming that the tree was grown according to growth model $\mathcal{M}$. We will call $P_{\mathcal{M}}(i | \mathcal{T})$ the \emph{root probability} of node $i$. (From here onwards we omit the subscript $\mathcal{M}$ from the root probability expression, always keeping in mind that there is an underlying growth model assumed.) We use the approach of \cite{shah2011rumors}, but here we formulate the problem directly as a set of message passing equations, that lead to a simple expression for the above probabilities, involving the nonbacktracking matrix of tree $\mathcal{T}$.

To calculate $P(i | \mathcal{T})$ we begin with Bayes' theorem, whereby

\begin{equation}
P(i | \mathcal{T}) = \frac{ P(\mathcal{T} | i) P(i) }{ P(\mathcal{T}) }.
\label{eq:3.1}
\end{equation}

\noindent
Here $P(\mathcal{T} | i)$, the ``likelihood'', is the probability that the result of the given generative process (growth model $\mathcal{M}$) is exactly tree $\mathcal{T}$, given that the root was node $i$.
The ``evidence'', $P(\mathcal{T}) = \sum_i^N P(\mathcal{T} | i) P(i)$, is independent of $i$ and serves only as a normalization constant, and hence need not be explicitly calculated. Assuming uniform prior probabilities $P(i)$, the root probabilities are simply proportional to the likelihood values, $P(i | \mathcal{T}) \propto P(\mathcal{T}|i)$. For a given growth model we can, in principal, find $P(\mathcal{T}|i)$ to solve our problem.
For growth models that have the EPS property (see Section \ref{sec2}) we can write

\begin{equation}
P(\mathcal{T} | i) = \pi_{\mathcal{T}} \mathcal{N}_{\mathcal{T},i} \propto P(i | \mathcal{T}),
\label{eq:3.2}
\end{equation}

\noindent
where $\mathcal{N}_{\mathcal{T},i}$ is the total number of different permitted link sequences started at node $i$ that result in tree $\mathcal{T}$, and $\pi_{\mathcal{T}}$ is the occurrance probability of any link sequence resulting in tree $\mathcal{T}$ under the given generative model. Thus, in a model that has the EPS property, the root probabilities of nodes are simply given by

\begin{equation}
P(i | \mathcal{T}) = \frac{ \mathcal{N}_{\mathcal{T},i} }{ \sum_{j=1}^N \mathcal{N}_{\mathcal{T},j}  },
\label{eq:3.3}
\end{equation}

\noindent
meaning that we only need to calculate $\mathcal{N}_{\mathcal{T},i}$ for each node $i$.

Let $\mathcal{B}_{i \leftarrow j}$ denote the set of nodes that can only be reached from $i$ via $j$ (including $j$), and let $\mathcal{N}_{\mathcal{B}_{i \leftarrow j}}$ be the number of different link sequences, starting with link $(i \leftarrow j)$, that result in branch $\mathcal{B}_{i \leftarrow j}$. With the help of all the $\mathcal{N}_{\mathcal{B}_{i \leftarrow j}}$ values incoming to node $i$, $\mathcal{N}_{\mathcal{T},i}$ may be expressed as

\begin{equation}
\mathcal{N}_{\mathcal{T},i} = \left[ \prod_{j \in \partial i} \mathcal{N}_{\mathcal{B}_{i \leftarrow j}} \right] \frac{ (N-1)! }{ \prod\limits_{j \in \partial i} N_{i \leftarrow j}! },
\label{eq:3.4}
\end{equation}

\noindent
with $N_{i \leftarrow j}$ denoting the number of nodes in branch $\mathcal{B}_{i \leftarrow j}$, including node $j$ but not $i$. The set of neighbours of node $i$ is denoted by $\partial i$.
The factor in square brackets gives the number of sequences resulting in tree $\mathcal{T}$, where the distinct branches $\mathcal{B}_{i \leftarrow j}$ emanating from node $i$ are built up consecutively. However, $\mathcal{N}_{\mathcal{T},i}$ is much greater than this number, since the branches $\mathcal{B}_{i \leftarrow j}$ in most cases will be built up simultaneously. The construction of the branches is independent in the sense that any link in the sequence generating branch $\mathcal{B}_{i \leftarrow j}$ may be exchanged with any link in the sequence generating $\mathcal{B}_{i \leftarrow k}$. The resulting complete sequence is still guaranteed to be permitted and still results in tree $\mathcal{T}$. The second factor on the right hand side of Eq. (\ref{eq:3.4}) gives just the number of ways this multiset permutation can be performed.
To proceed it will be useful to define

\begin{equation}
Q_i \equiv \frac{ \mathcal{N}_{\mathcal{T},i} } { (N-1)! },
\label{eq:3.5}
\end{equation}

\noindent
and

\begin{equation}
Q_{i \leftarrow j} \equiv \frac{ \mathcal{N}_{\mathcal{B}_{i \leftarrow j}} } { N_{i \leftarrow j}! }.
\label{eq:3.6}
\end{equation}

\noindent
Since $P(i | \mathcal{T}) \propto \mathcal{N}_{\mathcal{T},i}$, also $P(i | \mathcal{T}) \propto Q_i$ and we have that

\begin{equation}
P(i | \mathcal{T}) = \frac{ Q_i }{ \sum_{j=1}^N Q_j }.
\label{eq:3.7}
\end{equation}

\noindent
Using Eq. (\ref{eq:3.4}) we can also write

\begin{equation}
Q_i = \prod_{j \in \partial i} Q_{i \leftarrow j},
\label{eq:3.8}
\end{equation}

\noindent
meaning that it suffices to calculate the quantities $Q_{i \leftarrow j}$ in order to obtain the root probabilities $P(i | \mathcal{T})$.
We can establish a message passing (MP) type relationship between the $Q_{i \leftarrow j}$s by noticing that

\begin{equation}
\mathcal{N}_{B_{i \leftarrow j}} = \left[ \prod_{k \in \partial j \setminus i} \mathcal{N}_{\mathcal{B}_{j \leftarrow k}} \right] \frac{ (N_{i \leftarrow j}-1)! }{ \prod\limits_{k \in \partial j \setminus i} N_{j \leftarrow k}! },
\label{eq:3.9}
\end{equation}

\noindent
for the same reasons as above, in the case of Eq. (\ref{eq:3.4}).
Comparing Eqs. (\ref{eq:3.6}) and (\ref{eq:3.9}) we obtain

\begin{equation}
Q_{i \leftarrow j} = \frac{1}{ N_{i \leftarrow j} } \prod_{k \in \partial j \setminus i} Q_{j \leftarrow k}.
\label{eq:3.10}
\end{equation}

\noindent
Eq. (\ref{eq:3.10}) is a set of $2(N-1)$ equations, one for each directed link in tree $\mathcal{T}$. It is more convenient to solve the logarithm of Eq. (\ref{eq:3.10}),

\begin{equation}
\ln Q_{i \leftarrow j} = - \ln N_{i \leftarrow j} + \sum_{k \in \partial j \setminus i} \ln Q_{j \leftarrow k},
\label{eq:3.11}
\end{equation}

\noindent
which may be expressed in vector form,

\begin{equation}
\overrightarrow{\ln Q} = - \overrightarrow{\ln N} + \hat{B} \: \overrightarrow{\ln Q},
\label{eq:3.12}
\end{equation}

\noindent
where $\overrightarrow{\ln Q}$ and $\overrightarrow{\ln N}$ denote the element-wise logarithms of vectors $\vec{Q}$ and $\vec{N}$, respectively, and $\hat{B}$ is the nonbacktracking matrix of tree $\mathcal{T}$. The nonbacktracking matrix \cite{hashimoto1989zeta, krzakala2013spectral} is a matrix of $2L \times 2L$ entries (one row for each directed link), defined as $B_{i \leftarrow j, k \leftarrow l} = \delta_{jk}(1 - \delta_{il})$, where $i,j$ and $k,l$ are end node pairs of links in the graph.
The logarithmic formulation allows us to conveniently treat very large networks. (Note that the quantities $Q_{i \leftarrow j}$ are combinatorially large!)
The solution of Eq. (\ref{eq:3.12}) may be expressed as

\begin{equation}
\overrightarrow{\ln Q} = (\hat{B} - \hat{I})^{-1} \, \overrightarrow{\ln N},
\label{eq:3.13}
\end{equation}

\noindent
where $\hat{I}$ is the identity matrix.
The vector $\vec{N}$ can be easily calculated, since its elements are also related in a simple MP-type structure (recall that $N_{i \leftarrow j}$ is the number of nodes in branch $\mathcal{B}_{i \leftarrow j}$),

\begin{equation}
N_{i \leftarrow j} = 1 + \sum_{k \in \partial j \setminus i} N_{j \leftarrow k},
\label{eq:3.14}
\end{equation}

\noindent
or in vector form,

\begin{equation}
\vec{N} = \vec{1} + \hat{B} \, \vec{N},
\label{eq:3.15}
\end{equation}

\noindent
where $\vec{1}$ denotes the vector of all $1$s.
The solution of Eq. (\ref{eq:3.15}) is

\begin{equation}
\vec{N} = (\hat{I} - \hat{B})^{-1} \, \vec{1}.
\label{eq:3.16}
\end{equation}

For any given tree $\mathcal{T}$ and its nonbacktracking matrix $\hat{B}$ one can find $\vec{N}$ using Eq. (\ref{eq:3.16}), then plug this result into Eq. (\ref{eq:3.13}) to find $\overrightarrow{\ln Q}$, which then gives us the desired root probabilities via Eq. (\ref{eq:3.7}).

Although Eqs. (\ref{eq:3.13}) and (\ref{eq:3.16}) are compact and simple, it is generally not efficient numerically to obtain the solutions of Eqs. (\ref{eq:3.12}) and (\ref{eq:3.15}) by inverting the matrix $\hat{B} - \hat{I}$, as this operation has a time complexity worse than $O(n^2)$, where $n = 2(N-1)$ is the rank of matrix $\hat{B} - \hat{I}$ in general. It is much more efficient to obtain the solutions of Eqs. (\ref{eq:3.12}) and (\ref{eq:3.15}) iteratively, starting from any initial values. The solution (fixed point) of the equations will be obtained after at most $d_{\textrm{max}} - 1$ iterations, where $d_{\textrm{max}}$ is the maximum distance between any two nodes in tree $\mathcal{T}$, which behaves as $d_{\textrm{max}} \propto \ln N$ in ``small-world'' trees - the only type considered here. Hence the time complexity of solving Eqs. (\ref{eq:3.12}) and (\ref{eq:3.15}) iteratively is generally $O(N \ln N)$. It is possible, however, to obtain the solutions in linear time, at the expense of a slightly more complex implementation, by updating the message on each directed link exactly once.

In Section \ref{sec4} we will prove the EPS property (and hence, the validity of the above exact method for calculating the root probabilities) for general preferential attachment trees where the preference function is linear in node degree.

\subsection{Connection with closeness/distance centrality}

It is worth noting that the closeness and distance centrality on trees can be expressed using the nonbacktracking matrix similarly to Eqs. (\ref{eq:3.13}) and (\ref{eq:3.16}). Let us define the distance centrality of node $i$ as

\begin{equation}
D_i = - \sum_{j \neq i} d(i,j),
\label{eq:3.17}
\end{equation}

\noindent
where $d(i,j)$ is the length of the path between nodes $i$ and $j$. (The negative sign is used only for convenience.) The closeness centrality is usually defined as $C_i = -1/D_i$.
Let us write $D_i$ as a sum of incoming messages, $D_i = \sum_{j \in \partial i} D_{i \leftarrow j}$, where

\begin{equation}
D_{i \leftarrow j} = - \sum_{k \in \mathcal{B}_{i \leftarrow j}} d(i,k),
\label{eq:3.18}
\end{equation}

\noindent
so $D_{i \leftarrow j}$ is the ``distance centrality'' of directed link $i \leftarrow j$. These can be calculated using the simple recursive formula,

\begin{equation}
D_{i \leftarrow j} = - N_{i \leftarrow j} + \sum_{k \in \partial j \setminus i} D_{j \leftarrow k},
\label{eq:3.19}
\end{equation}

\noindent
or in vector form,

\begin{equation}
\vec{D} = - \vec{N} + \hat{B} \: \vec{D},
\label{eq:3.20}
\end{equation}

\noindent
with the solution

\begin{equation}
\vec{D} = (\hat{B} - \hat{I})^{-1} \, \vec{N} = - (\hat{B} - \hat{I})^{-2} \, \vec{1},
\label{eq:3.21}
\end{equation}

\noindent
where we used Eq. (\ref{eq:3.16}) in the last equality. Note the similarity with Eqs. (\ref{eq:3.13}) and (\ref{eq:3.16}).

\section{Range of applicability}
\label{sec4}

As discussed in Section \ref{sec3}, if a given tree $\mathcal{T}$ was grown according to a process that has the EPS property, then the root probability for all nodes can be calculated exactly using the method presented there. Due to simple symmetry reasons random recursive trees obviously have this property, as pointed out in Section \ref{sec2}. Here we show the EPS property for an important class of growth processes that is a generalization of the random recursive tree model.

We consider preferential attachment trees, where each new node connects to an existing node $i$ with probability proportional to $f(q_i)$, where $q_i$ is the degree of node $i$ and $f$ is a \emph{preference function}. Here we show that a preferential attachment tree growth model has the EPS property if the preference function has the form $f(q_i) = A + q_i$ with $A$ any real number in the range $[-1, \infty]$, or $f(q_i) = B - q_i$ with $B$ any integer number in the range $[2, \infty]$. We will refer to these growth models as positive and negative linear preferential attachment (LPA) models.

\subsection{Positive linear preferential attachment trees}
\label{sec41}

Let the preference function take the form

\begin{equation}
f(q_i) = A + q_i,
\label{eq:41.1}
\end{equation}

\noindent
where $A$ is an ``initial attractiveness'' of nodes, and is required to be $A \geq -1$. In the context of root finding, previous research has largely neglected this class of growing trees. It is, however, particularly important in network science, as it produces trees with degree distribution exponents in the empirically observed range $\gamma \in [2, \infty]$. To be precise, this growth model results in a scale-free degree distribution with exponent $\gamma = A + 3$. One can see that the random recursive tree model is a special case, when $A \rightarrow \infty$ and $\gamma \rightarrow \infty$. At the other extreme $A = -1$ produces a star.
For a given labeled tree $\mathcal{T}$ the occurrance probability of any link sequence that could have generated tree $\mathcal{T}$ can be written as

\begin{equation}
\pi_{\mathcal{T}} = \frac{ \prod\limits_{i=1}^N  \prod\limits_{j=1}^{q_i-1}  (A+j)  }{ \prod\limits_{t=3}^{N} \left[ A(t-1) + 2(t-2) \right]  },
\label{eq:41.2}
\end{equation}

\noindent
where $q_i$ is the degree of node $i$ in tree $\mathcal{T}$. This can be understood in the following way. In a given labeled tree $\mathcal{T}$, let us calculate the occurrance probability of a possible history (link sequence) $\mathcal{H} = \{(n_1 \leftarrow n_2), (p(n_3) \leftarrow n_3), \ldots, (p(n_N) \leftarrow n_N) \}$, where $n_t$ denotes the index of the node to arrive at step $t$, and $p(n_t)$ is the index of its parent node. The probability of the first link being realised is $1$, because all histories must start with a single link. In the second link (third time step) the new node has a choice of two equivalent existing nodes to attach to. This attachment, therefore, has a probability $1/2$. A subsequent attachment at time $t$ has probability $(A + q_{p(n_t)}(t-1)) / (\Theta(t-1))$, where $q_{p(n_t)}(t-1)$ denotes the degree of the parent node of node $n_t$ at time $t-1$, and $\Theta(t-1)$ is a normalization constant, which is simply $A$ times the number of nodes plus $2$ times the number of degrees in the network at time $t-1$, i.e., $\Theta(t-1) = A(t-1) + 2(t-2)$. Considering that after the first attachment (at time $t=2$) each node $i$ will serve as a ``parent'' exactly $q_i - 1$ times, it easily follows that the product of all the attachment probabilities is given by Eq. (\ref{eq:41.2}).
The probability $\pi_{\mathcal{T}}$ is clearly independent of the order in which the nodes were added to the tree.
Note that, although independent of node arrival order, $\pi_{\mathcal{T}}$ depends on tree $\mathcal{T}$ through its nodes' degrees.

\subsection{Negative linear preferential attachment trees}
\label{sec42}

Another type of linear preferential attachment model can be obtained by using the preference function

\begin{equation}
f(q_i) = \textrm{max}(B - q_i, 0),
\label{eq:42.1}
\end{equation}

\noindent
where the real parameter $B$ may be thought of as the ``capacity'' of a node: when $q_i > B$, node $i$ cannot attract any more nodes. The maximum degree in such a tree ensemble is the smallest integer number greater than or equal to $B$, and here lower degree nodes have a higher probability of attracting new nodes. A special case of this model is studied in \cite{shah2011rumors}, where $B$ is an integer number. This situation is equivalent to a slow SI spreading process inside a Bethe lattice of coordination number $B$, and is shown in \cite{shah2011rumors} to have the EPS property. This can be seen by writing, for a given labeled tree $\mathcal{T}$, the occurrance probability of any link sequence that could have generated tree $\mathcal{T}$,

\begin{equation}
\pi_{\mathcal{T}} = \frac{ \prod\limits_{i=1}^N  \prod\limits_{j=1}^{q_i-1}  (B-j)  }{ \prod\limits_{t=3}^{N} \left[ B(t-1) - 2(t-2) \right]  },
\label{eq:42.2}
\end{equation}

\noindent
which is independent of the order in which the nodes were added to the tree, but again depends on the degrees of nodes in the tree. (Equation (\ref{eq:42.2}) is similar to Eq. (\ref{eq:41.2}) and can be understood in an analogous manner.)
By letting $B \in [2, \infty]$ be a real number, we can continuously interpolate between random recursive trees ($B \to \infty$) and a chain ($B = 2$). For non-integer values of $B$, however, Eq. (\ref{eq:42.2}) no longer holds and the corresponding growth model does not have the EPS property.

\section{Inferring the root of growing scale-free trees}
\label{sec5}

For a tree $\mathcal{T}$ and given precision $\varepsilon$, a root finding algorithm \cite{bubeck2017finding} must return a set $K$ of nodes such that the actual root of the tree is a member of set $K$ with a probability at least $1-\varepsilon$, and $|K|$ must not depend on the system size. Bubeck et al. showed in \cite{bubeck2017finding} that root finding algorithms exist for random recursive and proportional preferential attachment trees. Here we show numerically that root finding is possible in the entire class of positive linear preferential attachment trees, and we assess its feasibility and limiting behaviour as we approach the two extremes: random recursive trees and a star.
The method described in Section \ref{sec3} is exact for positive LPA trees, therefore in these ensembles we obtain the optimal root finding algorithm by ranking all nodes according to their root probabilities and for any $\varepsilon$, the confidence set $K$ consists of the minimal set of nodes that have a cumulative root probability at least $1-\varepsilon$.
Our objective here is to find the functional form of the mean confidence set size $|K|$ versus desired inference precision $\varepsilon$, to assess the quality of root inference in the LPA class of trees.

\begin{figure}[t]
\centering
\includegraphics[width=\columnwidth,angle=0.]{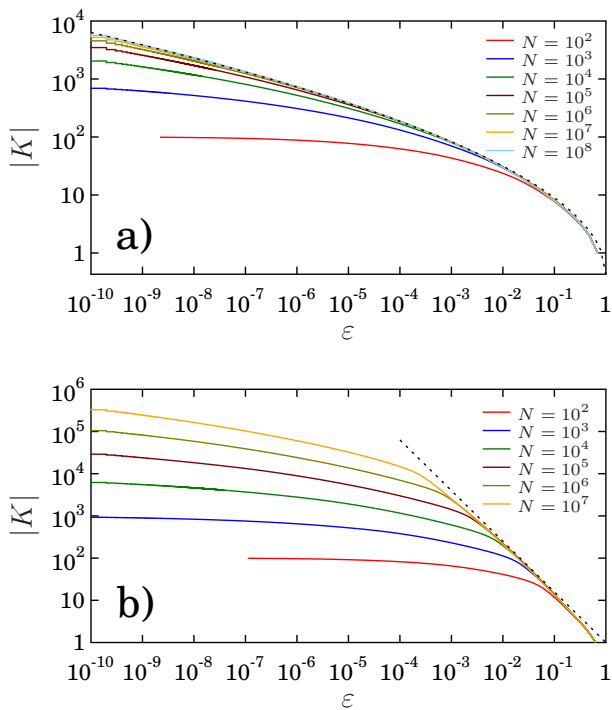}
\caption{(Color online) (a) Size $|K|$ of confidence set as a function of precision $\varepsilon$, averaged over different realizations of a random recursive tree, for various system sizes $N$. Dashed line corresponds to Eq. (\ref{eq:5.1}) with $c_1=0.43$, $c_2 = 2$. (b) $|K|$ as a function of $\varepsilon$ for proportinal preferential attachment trees. The function appears to converge to a power law. The fitted dashed line marks a power-law dependence with exponent $-1.2$. (The number of realizations in (a), (b) varied between $10^4$ for small sizes and $10$ for the biggest.)}
\label{fig:n_needed1}
\end{figure}

For random recursive trees, i.e., $A \to \infty$ (or $\gamma \to \infty$) Fig. \ref{fig:n_needed1}(a) shows $|K|$ as a function of $\varepsilon$ for different system sizes over $7$ orders of magnitude. Convergence to a size-independent function can be clearly seen. This function is well fitted by the form

\begin{equation}
|K| = c_1 e^{c_2 \sqrt{-\ln \varepsilon}},
\label{eq:5.1}
\end{equation}

\noindent
with parameters $c_1 = 0.43$, $c_2 = 2$. Eq. (\ref{eq:5.1}) agrees with the lower bound given in \cite{bubeck2017finding} to within a constant factor.

Figure \ref{fig:n_needed1}(b) shows $|K|$ as a function of $\varepsilon$ for proportional preferential attachment trees ($A = 0$ or $\gamma = 3$). Convergence to an asymptotic functional form is also apparent here. According to \cite{bubeck2017finding}, $|K|$ in this case is lower bounded by $c \varepsilon^{-1}$. Fig. \ref{fig:n_needed1}(b) indicates a power-law dependence with exponent close to $-1$, but the precise value cannot be obtained from the figure. Note the interesting finite size effect: at a certain size-dependent value of $\varepsilon$, $|K|$ appears to change abruptly from a power-law to a more slowly increasing function (with decreasing $\varepsilon$).

\begin{figure}[t]
\centering
\includegraphics[width=\columnwidth,angle=0.]{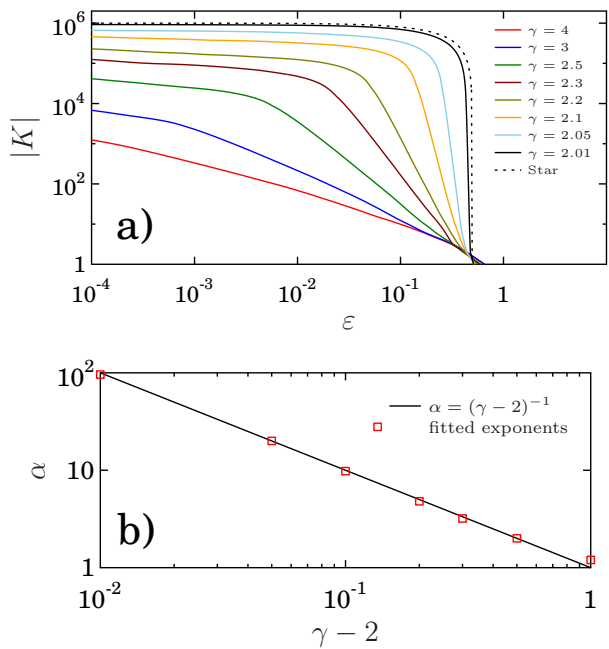}
\caption{(Color online) (a) $|K|$ as a function of $\varepsilon$ for positive linear preferential attachment trees of varying degree distribution exponent $\gamma$. The exact expression for $|K|$ for a star is plotted in dashed black line. The size of the tree was $N = 10^6$ in all cases and the curves were averaged over $10$ realizations. (b) Fitted exponents of the power-law behaviour seen in Fig. \ref{fig:n_needed2}(a).}
\label{fig:n_needed2}
\end{figure}

We can observe in Fig. \ref{fig:n_needed2}(a) that the power-law behaviour $|K| \propto \varepsilon ^{-\alpha}$ holds (up to a relatively well-defined transition point determined by system size) in general for positive LPA trees. The quality of the averaged curves allows for a relatively accurate fitting of the exponents. The exponents thus obtained are plotted in Fig. \ref{fig:n_needed2}(b). For $\gamma$ values close to $2$, the following dependence can be concluded,

\begin{equation}
\alpha \cong \frac{1}{\gamma - 2}.
\label{eq:5.2}
\end{equation}

\noindent
The fit becomes unreliable for larger values of $\gamma$. It may be hypothesized that for $N \to \infty$: $|K| \propto \varepsilon ^{-\alpha}$ with an $\alpha$ depending on $\gamma$, where $\alpha \in ]0, \infty]$. The nature of the crossover from this power-law behaviour to that described by Eq. (\ref{eq:5.1}) cannot be inferred from these studies.

\begin{figure}[t]
\centering
\includegraphics[width=\columnwidth,angle=0.]{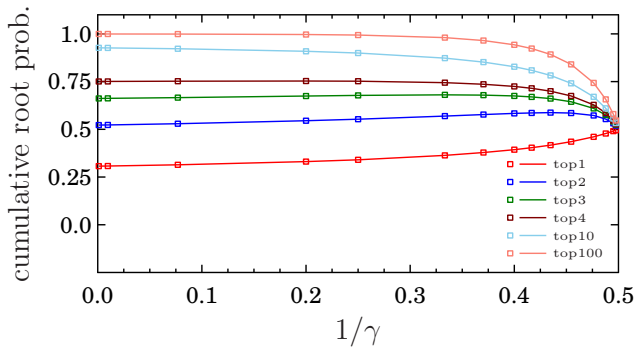}
\caption{(Color online) Cumulative root probability of top $n$ nodes in positive linear preferential attachment trees as a function of the inverse degree distribution exponent $1 / \gamma$. The cumulative root probabilities for all $n$ converge to $p = 1/2$ for $\gamma \to 2$. The tree size was $N = 10^4$ and the number of realizations $n = 10^5$ in all cases.}
\label{fig:rootprobs_plpa}
\end{figure}

\noindent
Interestingly, in the entire region $A \in ]-1, \infty]$ (or $\gamma \in ]2, \infty]$), $|K|$ appears to remain finite ($|K| \propto \varepsilon^{-\alpha}$) for any $\varepsilon$ as $N \to \infty$. However, when $A \to -1$ (or $\gamma \to 2$), the structure of the tree approaches a star, and in this case $|K| = N (1 - 2 \varepsilon)$ exactly, i.e., $|K|$ diverges as $N \to \infty$. The approach to this limiting behaviour can be seen in Fig. \ref{fig:n_needed2}(a), where $|K|$ for a star (of equivalent size) is plotted.
Note that the center of a star has root probability $1/2$ and all leaves of the star have root probability $1/[2(N-1)]$. Therefore, the formula $|K| = N (1 - 2 \varepsilon)$ is valid only in the range $\varepsilon \in [0, 1/2]$. For $\varepsilon > 1/2$, $|K| = 1$ in a star. In this limit the most probable root has the highest possible root probability, so up to an accuracy $1/2$ the root is easiest to infer in a star. With increasing $A$, as we move away from a star structure, the root probability of the most likely root decreases, so the quality of small accuracy root inference diminishes. On the other hand, root inference with high accuracy becomes possible, since $|K|$ does not diverge for any $\varepsilon$, and decreases monotonically with increasing $A$.
This phenomenon can also be verified in Fig. \ref{fig:rootprobs_plpa} where we plotted the cumulative root probabilities of the top $n$ nodes ($n = 1,2,3,4,10,100$) for positive LPA trees with varying $\gamma$. The cumulative probabilities of all ``top $n$'' sets of nodes converge to $1/2$ as $\gamma \to 2$. This means that as we approach this limit, the mean size $|K|$ of the confidence set for any given precision $\varepsilon < 1/2$, diverges, in accordance with Eq. (\ref{eq:5.2}).

\section{Reconstructing the complete history}
\label{sec6}

We have seen that the structure of a growing tree contains information about its past that allows for the determination of its most likely root with remarkable accuracy. Such information can be further utilized to make statements about the entire history of the given tree. One may be interested in the expected arrival times of individual nodes in the ensemble of all possible histories, which was studied recently in \cite{cantwell2019recovering}: an exact method was derived to calculate the mean arrival times in growth models that have the EPS property, and a Monte Carlo sampling approach to estimate it for general growth models. On the other hand one might wish to identify the single most likely history of a given growing tree. To consider the latter problem, in this section we extend the method of Section \ref{sec3} to estimate the most likely sequence of links (or nodes) that generated a given growing tree. As we saw earlier, due to the EPS property of the considered growth models, all link sequences that generate a given tree $\mathcal{T}$ have the same probability. Therefore, simply looking at the probability of a given complete sequence occurring, we cannot distinguish between them: neither is more likely than the others. However, the probabilities of partial sequences (say, the first 10 nodes) may be very different.
At any step of the growth process generating a given tree, the remaining nodes (or links) may have different probabilities of being the next one to be added. In this sense, we can talk about a ``stepwise most probable'' sequence: at each step we could calculate the probabilities of all remaining nodes to be the next added node, and always choose the one with the highest probability. It may be expected that this reconstruction has high accuracy in the beginning of the growth process, and the reconstruction is not likely to be much better than chance as we are approaching the periphery of tree $\mathcal{T}$.
We now discuss the method to find the stepwise most probable sequence that generated a given tree $\mathcal{T}$, assuming that the growth process had the EPS property. Our approach resembles the one applied in \cite{cantwell2019recovering}, essentially providing a faster ``greedy'' variant of the reconstruction strategy presented there.

\begin{figure}[t]
\centering
\includegraphics[width=6cm,angle=0.]{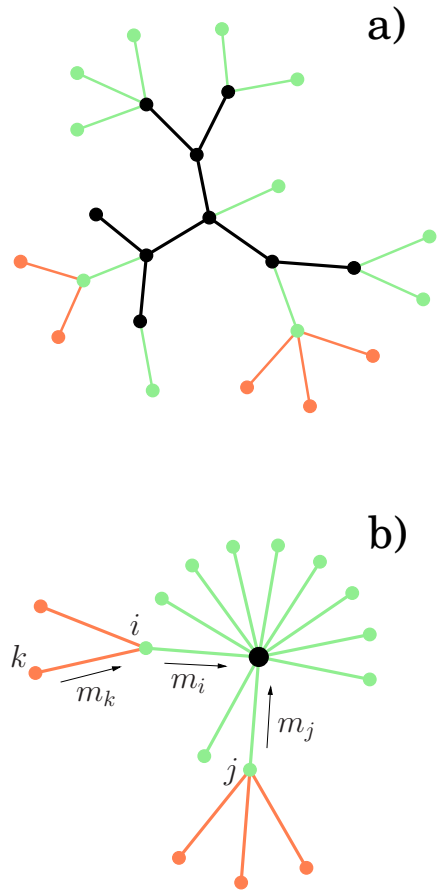}
\caption{(Color online). (a) A tree $\mathcal{T}$ consisting of $25$ nodes. The first $t=9$ nodes in the history of $\mathcal{T}$ are already known, and they constitute the subgraph $\mathcal{T}_{t=9} (\subset \mathcal{T})$, shown in black. The set of $11$ nodes that have nonzero probability of being the next added node, is shown in green, and the remaining $5$ nodes are shown in red. (b) ``Supernode'' representation of the tree in (a).}
\label{fig:tree_figure}
\end{figure}

Let us assume that we already know the first $t$ nodes in the history of tree $\mathcal{T}$. The first $t$ nodes, along with the links between them, form a subgraph of $\mathcal{T}$, let us denote this subgraph by $\mathcal{T}_t \subset \mathcal{T}$, see Fig. \ref{fig:tree_figure}(a). Our aim is, given $\mathcal{T}$ and $\mathcal{T}_t$, to calculate the probabilities $P(i|\mathcal{T}, \mathcal{T}_t)$ that node $i$ is the node added at time $t+1$. One can easily convince oneself, that if the EPS property holds for the complete tree $\mathcal{T}$ for a given growth model, then it also holds for the possible sequences that generate the remaining part of tree $\mathcal{T}$, after the first $t$ nodes in its history are already known. (It is easy to calculate $\pi_{\mathcal{T} \setminus \mathcal{T}_t}$, the occurrance probability of any sequence resulting in $\mathcal{T} \setminus \mathcal{T}_t$, to verify that this probability is independent of node arrival order.) Thus, for growth models that have the EPS property, we simply need to calculate the number of different sequences (of length $N-t$) that generate $\mathcal{T} \setminus \mathcal{T}_t$, starting at node $i$. Clearly there are no such sequences starting at nodes in $\mathcal{T} \setminus \mathcal{T}_t$ that are not directly adjacent to a node in $\mathcal{T}_t$. So at time $t+1$, we only need to consider - as viable candidates for being the next added node - neighbours of nodes in tree $\mathcal{T}_t$. We denote the set of such ``candidate'' nodes at a given step $t$ by $\mathcal{C}_t$.
Let us denote by $\mathcal{N}_{\mathcal{T} \setminus \mathcal{T}_t, i}$ the number of sequences starting at node $i$ that result in $\mathcal{T} \setminus \mathcal{T}_t$. The probability $P_{t+1}(i|\mathcal{T}, \mathcal{T}_t)$ that node $i$ was added at time $t+1$ can be written as

\begin{equation}
P_{t+1}(i | \mathcal{T}, \mathcal{T}_t) = \frac{ \mathcal{N}_{\mathcal{T} \setminus \mathcal{T}_t, i }}{ \sum_{j=1}^N \mathcal{N}_{\mathcal{T} \setminus \mathcal{T}_t, j}  }.
\label{eq:6.1}
\end{equation}

\noindent
To calculate $\mathcal{N}_{\mathcal{T} \setminus \mathcal{T}_t, i}$ it will be useful to define $N_{i \leftarrow j}$ as the number of nodes in branch $\mathcal{B}_{i \leftarrow j}$ that are not members of tree $\mathcal{T}_t$. We call such nodes ``free'' nodes, because at time $t$ they do not yet exist, and their birth time is not yet determined.
Similarly to Eq. (\ref{eq:3.4}) we can express $\mathcal{N}_{\mathcal{T} \setminus \mathcal{T}_t, i}$ as

\begin{equation}
\mathcal{N}_{\mathcal{T} \setminus \mathcal{T}_t, i} = \left[ \prod_{j \in \partial i} \mathcal{N}_{\mathcal{B}_{i \leftarrow j}} \right] \frac{ (N - 1 )! }{ \prod\limits_{j \in \partial i} N_{i \leftarrow j}! },
\label{eq:6.2}
\end{equation}

\noindent
where $N$ now denotes the number of free nodes in tree $\mathcal{T}$ at time $t$, i.e., $N = | \mathcal{T} \setminus \mathcal{T}_t | = N - t$. Eq. (\ref{eq:6.2}) may be seen simply as a generalization of Eq. (\ref{eq:3.4}). $\mathcal{N}_{\mathcal{B}_{i \leftarrow j}}$ now means the number of different sequences that result in the ``free part'' of branch $\mathcal{B}_{i \leftarrow j}$. (Note that the now introduced quantities $N$, $N_{i \leftarrow j}$ and $\mathcal{N}_{\mathcal{B}_{i \leftarrow j}}$ are generalizations of the quantities identified by the same notation in Section \ref{sec3}. This more general definition also applies there because in the root inference problem all nodes are ``free'' nodes.) Similarly to what we did before, let us define

\begin{equation}
Q_i \equiv \frac{ \mathcal{N}_{\mathcal{T} \setminus \mathcal{T}_t, i} } { (N - 1)! },
\label{eq:6.3}
\end{equation}

\noindent
and

\begin{equation}
Q_{i \leftarrow j} \equiv \frac{ \mathcal{N}_{\mathcal{B}_{i \leftarrow j}} } { N_{i \leftarrow j}! }.
\label{eq:6.4}
\end{equation}

\noindent
We can write

\begin{equation}
P_{t+1}(i | \mathcal{T}, \mathcal{T}_t) = \frac{ Q_i }{ \sum_{j=1}^N Q_j },
\label{eq:6.5}
\end{equation}

\noindent
and

\begin{equation}
Q_i = \prod_{j \in \partial i} Q_{i \leftarrow j}.
\label{eq:6.6}
\end{equation}

\noindent
As before, we can set up a MP-type relationship between the $Q_{i \leftarrow j}$s, by noticing that

\begin{equation}
\mathcal{N}_{B_{i \leftarrow j}} = \left[ \prod_{k \in \partial j \setminus i} \mathcal{N}_{\mathcal{B}_{j \leftarrow k}} \right] \frac{ (N_{i \leftarrow j} - f_{i \leftarrow j})! }{ \prod\limits_{k \in \partial j \setminus i} N_{j \leftarrow k}! }.
\label{eq:6.7}
\end{equation}

\noindent
Eq. (\ref{eq:6.7}) is identical to Eq. (\ref{eq:3.9}), except for the introduction of $f_{i \leftarrow j}$, which is a binary variable: $f_{i \leftarrow j} = 1$ if node $j$ is a free node, $f_{i \leftarrow j} = 0$ otherwise. This is so, because if node $j$ is not free, then the number of free nodes in branch $\mathcal{B}_{i \leftarrow j}$ does not decrease by one when we exclude node $j$. Note that $f_{i \leftarrow j}$ depends only on node $j$, and it is only for convenience that it is considered an attribute of link $(i \leftarrow j)$.
Using Eq. (\ref{eq:6.4}) we may write Eq. (\ref{eq:6.7}) as

\begin{equation}
Q_{i \leftarrow j} = N_{i \leftarrow j}^{-f_{i \leftarrow j}} \prod_{k \in \partial j \setminus i} Q_{j \leftarrow k}.
\label{eq:6.8}
\end{equation}

\noindent
Taking the logarithm of Eq. (\ref{eq:6.8}),

\begin{equation}
\ln Q_{i \leftarrow j} = - f_{i \leftarrow j} \ln N_{i \leftarrow j} + \sum_{k \in \partial j \setminus i} \ln Q_{j \leftarrow k},
\label{eq:6.9}
\end{equation}

\noindent
which may be expressed in vector form,

\begin{equation}
\overrightarrow{\ln Q} = - \vec{f} \circ \overrightarrow{\ln N} + \hat{B} \: \overrightarrow{\ln Q},
\label{eq:6.10}
\end{equation}

\noindent
where $\circ$ denotes the Hadamard product.
The solution may be expressed as

\begin{equation}
\overrightarrow{\ln Q} = (\hat{B} - \hat{I})^{-1} \, (\vec{f} \circ \overrightarrow{\ln N}).
\label{eq:6.11}
\end{equation}

\noindent
The vector $\vec{N}$ can be easily calculated, since its elements are also related in a simple MP-type structure (recall that $N_{i \leftarrow j}$ is the number of free nodes in branch $\mathcal{B}_{i \leftarrow j}$),

\begin{equation}
N_{i \leftarrow j} = f_{i \leftarrow j} + \sum_{k \in \partial j \setminus i} N_{j \leftarrow k},
\label{eq:6.12}
\end{equation}

\noindent
or in vector form,

\begin{equation}
\vec{N} = \vec{f} + \hat{B} \, \vec{N}.
\label{eq:6.13}
\end{equation}

\noindent
The solution of Eq. (\ref{eq:6.13}) is

\begin{equation}
\vec{N} = (\hat{I} - \hat{B})^{-1} \, \vec{f}.
\label{eq:6.14}
\end{equation}

To construct the stepwise most probable sequence resulting in tree $\mathcal{T}$, at each step we may solve Eqs. (\ref{eq:6.13}) and (\ref{eq:6.10}) iteratively, and use Eqs. (\ref{eq:6.6}) and (\ref{eq:6.5}) to find the probability for each remaining node of being the next added node. We then choose the node with the highest probability and add it to tree $\mathcal{T}_t$ to obtain tree $\mathcal{T}_{t+1}$. This sequence is the exact stepwise most probable sequence to result in tree $\mathcal{T}$ assuming that tree $\mathcal{T}$ was grown according to a growth model that has the EPS property.
Note that using the above method we can calculate the occurrance probability of any partial sequence, not just the most probable one!

\subsection{Linear-time implementation}
\label{sec61}

The above algorithm can be greatly simplified by noticing some simple properties of the structure of the MP equations.
Looking at Eq. (\ref{eq:6.9}) we notice that whenever node $j$ is a non-free node, the message $\ln Q_{i \leftarrow j}$ is calculated simply as the sum of the messages incoming to directed link $(i \leftarrow j)$. In this sense, only free nodes ``add complexity'', and non-free nodes serve merely as ``summing junctions''. This means that the entire existing part, $\mathcal{T}_t$, could be collapsed into one non-free supernode, and the resulting solutions would remain the same for all $\ln Q_{i \leftarrow j}$ where at least one of the nodes $i$, $j$ is free. Figure \ref{fig:tree_figure}(b) shows this ``supernode'' representation of the tree in Fig. \ref{fig:tree_figure}(a).
We can make further simplifications by defining the ``upstream'' messages of nodes. For a free node $i$ let $m_i \equiv \ln Q_{i' \leftarrow i}$ where $i'$ is the ``upstream'' neighbour of $i$, i.e., the neighbour that is closest to the existing tree $\mathcal{T}_t$ (or the supernode), see Fig. \ref{fig:tree_figure}(b). We denote this ``upstream'' neighbour of node $i$ by $u(i)$. With these definitions we can express $\ln Q_i$ for all candidate nodes $i \in \mathcal{C}$ using Eq. (\ref{eq:6.6}), Eq. (\ref{eq:6.9}) and using the fact that the supernode is only a summing junction:

\begin{equation}
\ln Q_i = \sum_{j \in \mathcal{C}} m_j - m_i + \sum_{k \in \partial_i \setminus u(i)} m_k.
\label{eq:6.15}
\end{equation}

\noindent
Also, using Eq. (\ref{eq:6.9}) we can write

\begin{equation}
m_i = -\ln N_i + \sum_{k \in \partial_i \setminus u(i)} m_k,
\label{eq:6.16}
\end{equation}

\noindent
where we introduced the ``downstream'' branch (DSB) size of node $i$ as $N_i \equiv N_{u(i) \leftarrow i}$.
Comparing Eqs. (\ref{eq:6.15}) and (\ref{eq:6.16}) we obtain

\begin{equation}
\ln Q_i = \sum_{j \in \mathcal{C}} m_j + \ln N_i.
\label{eq:6.17}
\end{equation}

\noindent
Exponentiating, we get

\begin{equation}
Q_i = N_i \, e^{\sum_{j \in \mathcal{C}} m_j}.
\label{eq:6.18}
\end{equation}

\noindent
We see that $Q_i$ depends on $i$ only through $N_i$, the DSB size of node $i$.
Remembering Eqs. (\ref{eq:6.1}) and (\ref{eq:6.3}) we finally have

\begin{equation}
P_i \equiv P_{t+1}(i | \mathcal{T}, \mathcal{T}_t) \propto N_i,
\label{eq:6.19}
\end{equation}

\noindent
in other words, the probability that node $i$ is the next node is proportional to the downstream branch size of node $i$.
Based on this result, the probability of any partial sequence can be calculated in time linear in $\ell$, the length of the sequence. Furthermore, the ensemble of possible histories on a given tree can be uniformly sampled by the following procedure. First choose a node at random, as the root, with probability calculated in Eq. (\ref{eq:3.7}). Then, in each step, add a node from the periphery chosen with probability proportional to its downstream branch size. As discussed in \cite{cantwell2019recovering} this procedure may also be applied to growth models that do not have the EPS property: the Monte Carlo samples must be weighted by the occurrance probability of the given sequence under the specific growth model.

\subsection{Fast history reconstruction}
\label{sec62}

Using the above results we propose the following algorithm to reconstruct the complete history of a growing tree $\mathcal{T}$.
First we use the root inference algorithm of Section \ref{sec3} to find the most probable root of $\mathcal{T}$. Then we perform a search from this node to find the DSB size $N_i$ of all nodes $i$ in $\mathcal{T}$. Our reconstruction algorithm then simply consists of sorting the nodes of $\mathcal{T}$ in decreasing order of $N_i$ to obtain the estimated sequence of node arrivals, which is exactly the stepwise most probable sequence in the case of growth models with the EPS property.

\subsubsection{Dealing with symmetries}
\label{sec621}

It is immediately obvious that at a given step $t$ all leaves adjacent to $\mathcal{T}_t$ have the same DSB size $ = 1$, and are, therefore, equally probable to be the next added node. That is, in terms of probability at step $t$, there is no way to distinguish between them. Such symmetry classes also exist for higher order structures, not just leaves.
A simple empirical method (similar to what is done in \cite{sreedharan2019inferring}) to distinguish between nodes of equal DSB size, is to consider the average DSB size of their neighbours. The reasoning for this is that if the DSB size is an indicator of age, then nodes with on-average older neighbours aught to be older than other nodes of equal DSB size.
Therefore, to each node $i$ in $\mathcal{T}$ we assign the number

\begin{equation}
H_i = N_i + \frac{1}{N q_i} \sum_{j \in \partial i} N_j,
\label{eq:621.1}
\end{equation}

\noindent
where $q_i$ is the degree of node $i$ and $N$ is the total number of nodes in tree $\mathcal{T}$. We then sort all nodes $i$ in decreasing order of $H_i$ to arrive at the estimated sequence of node arrivals. (In Eq. (\ref{eq:621.1}) we divided the average DSB size of neighbours by $N$ to ensure that this additional term is less than $1$, i.e., does not change the estimated arrival order of nodes that have different DSB sizes.) We will refer to this method as \texttt{SMPH} (Stepwise Most Probable History).

\subsection{Other fast reconstruction methods}
\label{sec63}

Here we give a list of linear-time, or almost linear-time, tree history reconstruction methods that we compare with the \texttt{SMPH} method described in Section \ref{sec621}. (When assessing the time complexity of a reconstruction method, we only consider the time required to calculate certain desired quantities for all nodes according to which they can be ranked. We do not consider the actual sorting of the nodes, which has a worst-case time complexity of $O(N \ln N)$.) As a benchmark, we consider a \texttt{Stepwise Random} reconstruction. This means choosing a root node uniformly at random, and then adding nodes sequentially from the existing periphery always uniformly at random. The method \texttt{High Degree} chooses the highest degree node as root, and always adds the highest degree node from the existing periphery, ties broken uniformly at random. A similar degree-based method is \texttt{Low Degree}, that reconstructs the history in reverse, always choosing the lowest degree node in every step, ties again broken at random. (This is essentially a random pruning process.) A more sophisticated reverse reconstruction method, \texttt{Peeling+}, introduced in \cite{sreedharan2019inferring}, first assigns a layer index to nodes by sequentially removing all leaves from the tree at once in each step. After all nodes have been assigned layer indices, a reconstructed reverse node sequence is given by ordering nodes according to their layer indices, starting with nodes that were initially leaves. Ties are broken by considering the mean layer index of the neighbours of nodes. In addition to these centrality-based methods, we also study history reconstruction based on the results recently presented in \cite{cantwell2019recovering}. A method was introduced to exactly calculate the mean arrival times of nodes in a tree, averaged over all possible histories of the given tree, assuming a generative process that has the EPS property. The algorithm, however, runs in quadratic time (in system size $N$), therefore is only a viable option for relatively small networks. The mean arrival times can be estimated by Monte Carlo sampling, however, as described in Section \ref{sec61} and \cite{cantwell2019recovering}.

\section{Numerical results}
\label{sec7}

In this section we present numerical results of tree history reconstruction using the \texttt{SMPH} method described in Section \ref{sec621}. We compare the reconstruction quality with other existing methods that run in linear time or almost linear time. In Section \ref{sec72} we use our logarithmic MP method to calculate mean logarithmic history degeneracies in various growing tree models, in other words, the mean logarithm of the number of histories concordant with a given static tree structure produced by a given growth model. Due to the high precision of the measurements we can arrive at some reliable general conclusions.

\subsection{History reconstruction of growing trees}
\label{sec71}

We compare our history inference method (based on Eq. (\ref{eq:621.1})) with other linear-time algorithms using a global measure of reconstruction quality: the average Pearson rank correlation coefficient between the ranking of nodes in the real history and the inferred one. Let $x_i$ be the rank of node $i$ in the true arrival order and let $\tilde{x}_i$ be the rank of node $i$ in the inferred sequence. Then the Pearson rank correlation coefficient is defined as

\begin{equation}
\rho = \frac{\textrm{cov}(x,\tilde{x})}{\sigma_x \sigma_{\tilde{x}}},
\label{eq:71.1}
\end{equation}

\noindent
where $\textrm{cov}$ denotes covariance and $\sigma$ is the standard deviation. The coefficient $\rho$ can take values in the range $[-1, 1]$. Identical sequences produce $\rho = 1$, completely uncorrelated sequences produce $\rho = 0$, and inverted sequences result in $\rho = -1$.

\begin{figure}[H]
\centering
\includegraphics[width=\columnwidth,angle=0.]{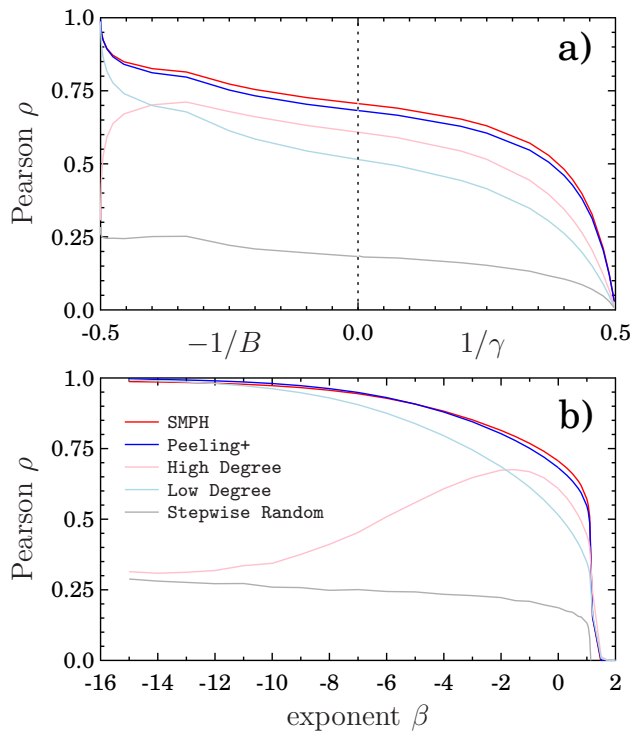}
\caption{(Color online). Pearson rank correlation coefficient for the linear-time history reconstruction algorithms considered, for (a) linear and (b) nonlinear preferential attachment trees in the complete range between a chain ($B \to 2$, $\beta \to -\infty$) and a star ($\gamma \to 2$, $\beta \to 2$). Dashed vertical line in (a) represents random recursive trees ($A, B \to \infty$, $\gamma \to \infty$). Tree size in all cases was $N=10^5$. Results were averaged over at least $100$ realizations.}
\label{fig:Pearson}
\end{figure}

Figure \ref{fig:Pearson}(a) presents Pearson coefficients for our history inference method on linear preferential attachment trees, compared with the other linear-time methods discussed in Section \ref{sec63}. Using $-1/B$ and $1/\gamma$ as parameters, it is possible to present the complete range between the chain and star limits in one continuous plot. (Note the smooth transition at $-1/B = 1/\gamma = 0$.)

The \texttt{SMPH} method performs the best in almost the entire range, except very close to the \emph{chain} limit ($-1/B \to -0.5$). In the \emph{star} limit no inference is possible, since all nodes (except the central node) are structurally equivalent. In this limit $\rho$ for all methods necessarily goes to zero. In the \emph{chain} limit the structure of the tree contains maximal information and $\rho$ approaches $1$ for all methods except \texttt{High Degree}, which converges to the inference quality of a stepwise random sequence. Note that the \texttt{Stepwise Random} method still has positive reconstruction quality.
Our method performs slightly better than \texttt{Peeling+}, which is already close to optimal in proportional preferential attachment graphs, as demonstrated in \cite{sreedharan2019inferring}.

We tested the \texttt{SMPH} algorithm also on nonlinear preferential attachment trees (NPA), where the preference function is a power of node degree,

\begin{equation}
f(q) = q^{\beta}.
\label{eq:71.2}
\end{equation}

\noindent
This class of growth model has also received some attention due to the very different structures it can produce \cite{krapivsky2000connectivity}. A negative exponent promotes low degree nodes as hosts for new attachment, resulting in elongated, thin branches and very rapidly decaying degree distribution. Such a process has been used to model discussion cascades in social media \cite{gomez2011modeling}. Positive exponents produce wider degree distributions, most notably $\beta = 1$ results in a scale-free degree distribution with exponent $-3$. For $\beta > 1$, the model enters a \emph{condensation} regime, where a single node has a finite fraction of all degrees in the network. The condensation is complete at $\beta = 2$ where the structure of the tree is essentially a star \cite{young2019phase}.

Figure \ref{fig:Pearson}(b) shows Pearson coefficients for the inference methods considered, on NPA trees. A similar trend to that of Fig. \ref{fig:Pearson}(a) can be seen, although the approach to the chain limit is more ``stretched out''. Apart from the extreme negative exponent range, our method performs the best, again slightly better than \emph{Peeling+}. For positive exponents, the inference quality of all methods goes to zero somewhere in the range $1 < \beta < 2$, in accordance with \cite{young2019phase}, where it is suggested that an inferrability phase transition takes place for a certain $\beta_c \in [1, 2]$.

\begin{figure}[t]
\centering
\includegraphics[width=\columnwidth,angle=0.]{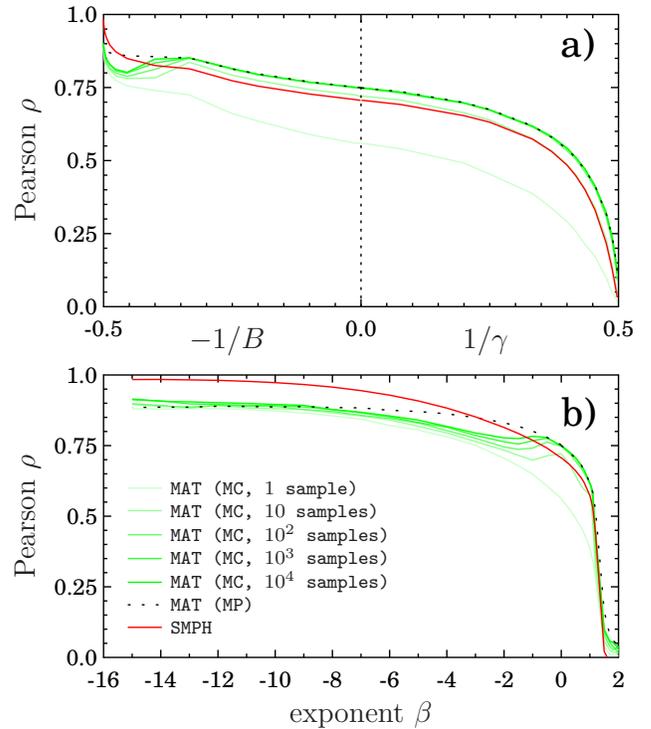}
\caption{(Color online). Pearson correlation coefficient for mean arrival time (\texttt{MAT}) values, calculated according to the MP method of \cite{cantwell2019recovering} and estimated via Monte Carlo (MC) sampling. The \texttt{SMPH} method is shown for comparison. Growth models considered were (a) linear and (b) nonlinear preferential attachment, for the same parameters as in Fig. \ref{fig:Pearson}. Tree size in all cases was $N=10^3$. Results were averaged over $100$ realizations of the tree structure. \texttt{MAT}-based methods perform slightly better than the \texttt{SMPH} method for more compact trees with a large fraction of leaf nodes, while the \texttt{SMPH} method is more efficient for more chain-like trees with a smaller fraction of leaf nodes (smaller values of $-1/B$ and $\beta$).}
\label{fig:Pearson_MC}
\end{figure}

We also investigated the reconstruction quality of the more sophisticated, albeit more time-consuming methods presented in \cite{cantwell2019recovering}. A message-passing (MP) method was developed to calculate the exact mean arrival times of nodes in a tree, averaged over all possible histories of the given tree, assuming a generative process that has the EPS property. For tree generative models that do not have the EPS property, a Monte Carlo (MC) sampling method may be used to estimate the exact mean arrival times, provided that the probability of a given history can be evaluated under the given generative model (see \cite{cantwell2019recovering}). It was shown in \cite{young2019phase} that the exact mean arrival times are optimally correlated with the actual history of a given tree, i.e. have the highest Pearson coefficient of all possible methods to estimate the history.
However, the MP method runs in quadratic time (in system size $N$), therefore is not suitable for large trees. The runtime is also proportional to the mean ``excess degree'', $(\langle k^2 \rangle - \langle k \rangle) / \langle k \rangle$, which further limits the applicability of the method for trees with broad degree distributions.
One Monte Carlo sample of a potential history can be produced in almost linear time (see \cite{cantwell2019recovering}), but the typical number of samples required for a good approximation of the exact mean arrival times of nodes is not known, and may vary significantly for different generative models (different types of tree structures).

Considering trees of size $N = 10^3$, Figure \ref{fig:Pearson_MC}(a) confirms that $100$ samples are typically enough for reliable convergence for linear preferential attachment models, except in the range close to the chain limit. The Pearson correlation for the mean arrival times (\texttt{MAT}) computed according to the MP method coincide with the Pearson coefficients for the \texttt{MAT} based on MC samples (when fully converged) in the range when the generative model has the EPS property. In this range the Pearson coefficients for the \texttt{MAT} (which are optimal) are slightly higher than that for the \texttt{SMPH} method. Close to the chain limit the MP method does not return the exact mean arrival times---the EPS property is not fulfilled---, and is therefore not optimal. Given enough samples the MC method should still converge to the optimal estimate, but, as Fig. \ref{fig:Pearson_MC}(a) shows, convergence slows down considerably in this range. The reason for this is that in this case the probabilities of different possible histories vary considerably more than in ``more compact'' trees, and only a very small fraction of MC samples carry significant weight in the average.
This effect is even more apparent in Fig. \ref{fig:Pearson_MC}(b), where it is clearly seen that convergence of the MC method is prohibitively slow for more chain-like trees, i.e., trees with a small fraction of leaf nodes. In such cases, where the exact \texttt{MAT} values cannot be estimated in practice, the \texttt{SMPH} method provides a good alternative for high quality history inference.

\subsection{History degeneracy of growing trees}
\label{sec72}

The fundamental reason for the difficulty of history inference is that any given tree structure may have been produced by a very large number of different histories. Let us define the \emph{history degeneracy} $\mathcal{N}_{\mathcal{T}}$ of a tree $\mathcal{T}$ as the number of different link sequences that may have produced tree $\mathcal{T}$. This number is related with the information content \cite{rashevsky1955life, trucco1956note} and the structure of the automorphism group of the given tree. These concepts also play a central role in graph structure compression. A chain has maximal information content and maximal history inferrability, as we saw in Section \ref{sec71}. This translates to a minimal history degeneracy. The other limit, a star, has minimal information content, minimal history inferrability and maximal history degeneracy.
In these limits $\mathcal{N}$ can be easily calculated for a tree of $N$ nodes,

\begin{align}
\ln \mathcal{N}_{\textrm{chain}}(N) &= \ln \left[ \sum_{i=1}^N \binom{N-1}{i-1} \right] = \ln (2^{N-1}) \cong N \ln 2, \label{eq:72.1}\\
\ln \mathcal{N}_{\textrm{star}}(N) &= \ln[2(N-1)!] \cong N \ln N - N \label{eq:72.2}.
\end{align}

\noindent
(As $\mathcal{N}$ is generally very large, it is more convenient to work with its logarithm.)
Since $\ln \mathcal{N}_{\textrm{chain}}$ and $\ln \mathcal{N}_{\textrm{star}}$ have qualitatively different dependence on $N$, it is interesting to consider how $\ln \mathcal{N}$ behaves between these two limiting structures.

\begin{figure}[H]
\centering
\includegraphics[width=\columnwidth,angle=0.]{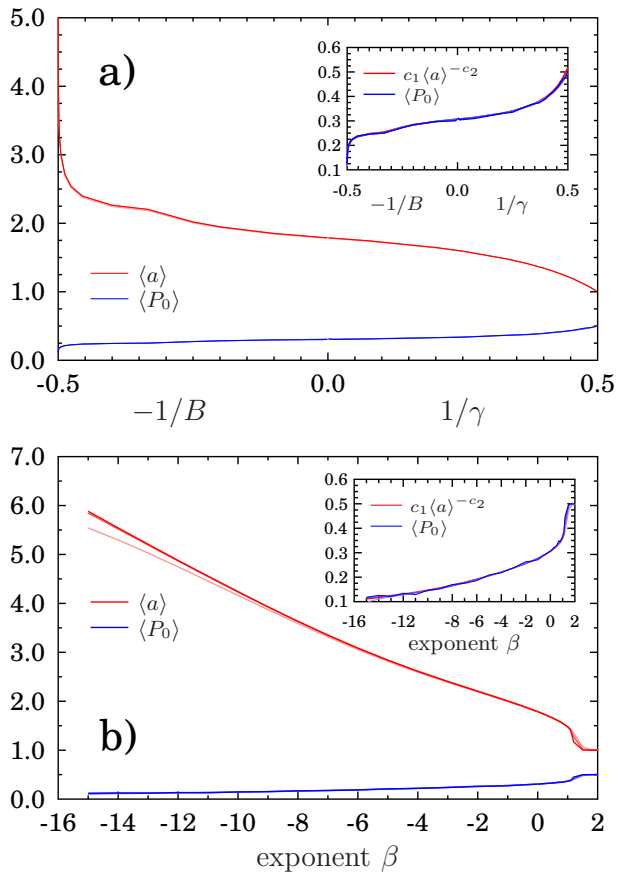}
\caption{(Color online). Averaged $\langle a \rangle$ and $\langle P_0 \rangle$ values for the entire range between the chain and the star limit, for (a) linear and (b) nonlinear preferential attachment trees. Curves plotted correspond to sizes $N = 10^3$, $10^4$ and $10^5$, in different shades of red ($\langle a \rangle$) and different shades of blue ($\langle P_0 \rangle$). In all cases, curves for different sizes overlap almost perfectly. Insets present rescaled values $c_1 \langle a \rangle^{-c_2}$, which coincide almost perfectly with $\langle P_0 \rangle$. The constants for the two model classes were: (a) $c_1 = 0.52$, $c_2 = 0.91$ and (b) $c_1 = 0.5$, $c_2 = 0.85$. Results were averaged over at least $100$ realizations.}
\label{fig:a_P0}
\end{figure}

The number of possible histories of a tree is a quantity that plays a central role in the inference approaches discussed in \cite{shah2011rumors, bubeck2017finding, cantwell2019recovering}. However, its size dependence (for a given generative model) has not been investigated in these works.
The logarithmic MP scheme presented in Section \ref{sec3} enables us to conveniently calculate $\ln \mathcal{N}$ for any given tree, as

\begin{align}
\ln \mathcal{N} &= \ln (N-1)! + \ln \sum_{i=1}^N Q_i\\
&= \ln (N-1)! + \ln Q_0 - \ln P_0\\
&\cong N \ln N - N + \ln Q_0,
\label{eq:72.3}
\end{align}

\noindent
where $Q_0 = \textrm{max}(Q_i)$, $P_0 = \textrm{max}(P(i|\mathcal{T}))$, and we used Eqs. (\ref{eq:3.5}) and (\ref{eq:3.7}). Note that only logarithmic quantities are involved, enabling us to consider large networks.
We calculated $\ln \mathcal{N}$ in this way for the full range of LPA and NPA trees between the chain and star limit, for different tree sizes $N$. In Fig. \ref{fig:a_P0} we plotted the quantity

\begin{equation}
a = \frac{ N \ln N - \ln \mathcal{N} }{ N } \left( \cong 1 - \frac{\ln Q_0}{N} \right),
\label{eq:72.4}
\end{equation}

\noindent
averaged over the ensemble of trees defined by a given parameter set. (The number ``$a$'' is a random variable that has some distribution in the ensemble of trees generated by a given growth model.)
The figure confirms that $\langle a \rangle$ rapidly converges to a constant with increasing $N$ everywhere between the star and chain limit, implying the following general functional form for history degeneracy,

\begin{equation}
\langle \ln \mathcal{N} \rangle \cong N \ln N - cN,
\label{eq:72.5}
\end{equation}

\noindent
where $c = \langle a \rangle$ is independent of size and depends only on model parameters. Clearly, $c \to 1$ as we approach a star, and $c \to \infty$ as we approach a chain, as is confirmed by Fig. \ref{fig:a_P0}. This divergence appears to be very slow, but the exact form cannot be concluded from these studies. Magner et al. show in \cite{magner2017recovery} that for proportional preferential attachment (i.e., positive LPA with $A = 0$) $\langle \ln \mathcal{N} \rangle = (1+o(1))N \ln N$ and $\langle \ln \mathcal{N} \rangle \geq N \ln N + O(N)$. However, to our knowledge, the functional form of Eq. (\ref{eq:72.5}) has not yet been strictly shown for any growing tree model.

Additionally, for the two model classes considered, we calculated $\langle P_0 \rangle$, the mean root probability of the most probable root node (assuming that the tree was grown according to a model that has the EPS property). This quantity may be calculated for any tree generative model, but only gives the exact mean root probability for growth models with the EPS property. (Note that the LPA model has this property for any value of the parameter $A$ and integer values of $B$, but the NPA model only has it for $\beta=0$ and $\beta=1$) For all models considered, we found an interesting approximate scaling relationship between $\langle a \rangle$ and $\langle P_0 \rangle$,

\begin{equation}
\langle P_0 \rangle = c_1 \langle a \rangle^{-c_2},
\label{eq:72.6}
\end{equation}

\noindent
where $c_1$ and $c_2$ are positive constants for a given model class (see the insets of Fig. \ref{fig:a_P0}). This form appears to hold in almost the entire range between the chain and star limits, for LPA and NPA trees with high accuracy. The constants for the two different model classes can be found in the caption of Fig. \ref{fig:a_P0}.
Equation (\ref{eq:72.6}) demonstrates an interesting ``uncertainty principle'' for history inference in growing trees: the higher the inferrability of the complete history (higher value of $\langle a \rangle$), the lower the inferrability of the root.

Further numerical results regarding the statistics of $P_0$ and $a$ are presented in Appendix \ref{app2}.

\section{Conclusions}
\label{sec8}

In this paper we have extended on existing exact solutions to the root finding problem on growing trees to investigate the possibility of inferring the sequence of node arrivals in a broad class of recursive trees. We have formulated the root finding problem as a set of message passing equations which are exact for tree growth models where all possible histories of a given tree occur with the same probability. We have shown that this is true for growing preferential attachment trees with a linear preference function. Based on the message passing formulation we have proposed a simple algorithm to generate the stepwise most probable sequence of node arrivals of a given tree. In terms of computational complexity the method is optimal, i.e., runs in linear time.

We have compared the reconstruction quality of our method to existing linear-time inference algorithms for two broad classes of growing tree models: linear and nonlinear preferential attachment trees. Both of these model classes interpolate between a chain and a star structure, and both contain random recursive trees and proportional preferential attachment as special cases.
Compared with other linear-time algorithms our method performs the best in almost the entire range. It is almost as good as the more time-consuming optimal method based on the mean arrival times of nodes in the ensemble of all possible histories of the given tree. The exact calculation of mean arrival times can only be done in quadratic time, and only for growth models that produce equiprobable histories. A universally applicable Monte Carlo estimate of the mean arrival times, although more efficient in most cases, also runs in super-linear time. Furthermore, convergence in the Monte Carlo strategy becomes prohibitively slow for trees that have a structure close to a chain (small fraction of leaf nodes). In such networks, and in large networks in general, our method may be the preferred option.

Our message passing formulation also allows us to calculate the mean logarithmic history degeneracy of large trees, and make general observations for a broad class of tree growth models. We have found the general formula $\langle \ln \mathcal{N} \rangle \cong N \ln N - cN$ for the mean logarithmic history degeneracy in terms of system size, which is valid for all linear and nonlinear preferential attachment growth models with very high precision. We have also found a scaling relationship between the mean logarithmic history degeneracy and the mean root probability of the most likely root, in the growth model classes considered. This ``uncertainty principle'' highlights an interesting tradeoff inherent in the reconstruction problem: the root and the complete history of a tree cannot be inferred with high accuracy at the same time.

Although only applicable to trees directly, our results may also have implications for history inference on loopy networks. For example, the network of contacts between infected individuals in an epidemic may contain many loops, but the actual disease transmissions must have followed the links of a spanning tree of the network. Information about the probability distribution over the ensemble of spanning trees, under a given epidemic model, would create a bridge between tree history inference methods and history inference on general, loopy networks. This is an interesting challenge for the future.

\section*{Acknowledgments}

We thank Jean-Gabriel Young, George T. Cantwell and Guillaume St-Onge for useful comments.
This work was developed within the scope of the project i3N, UIDB/50025/2020 \& UIDP/50025/2020, financed by national funds through the FCT/MEC--Portuguese Foundation for Science and Technology. G. T. and R. A. C. were supported by FCT grants CEECIND/03838/2017 and CEECIND/04697/2017.

\appendix

\section{Statistics of $P_0$ and $a$}
\label{app2}

Here we study the statistics of $P_0$ (the root probability of the most likely root) and the number $a$ (defined in Eq. (\ref{eq:72.4})) numerically.

\begin{figure}[t]
\centering
\includegraphics[width=\columnwidth,angle=0.]{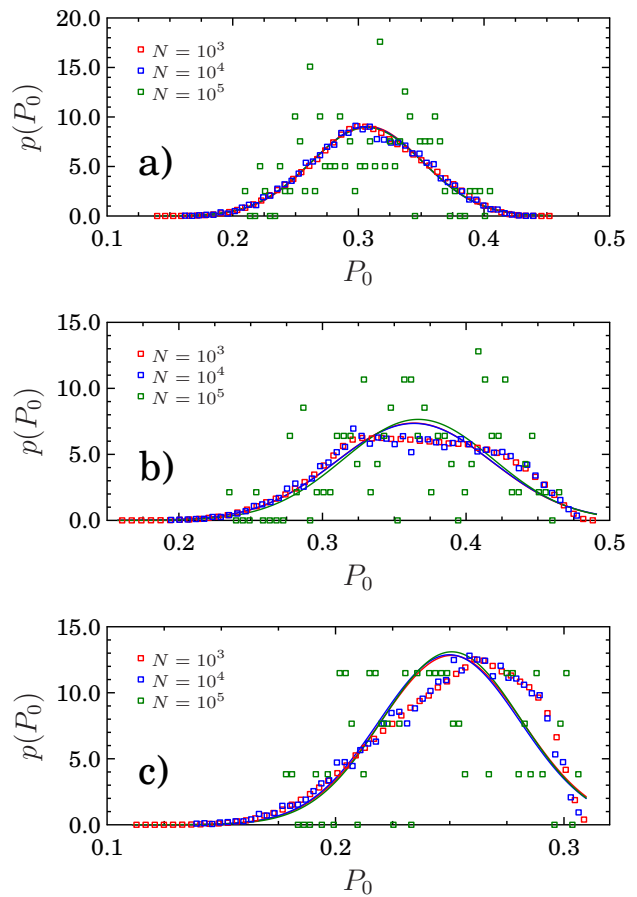}
\caption{(Color online). Distribution of $P_0$ values for three sizes, $N = 10^3$, $10^4$ and $10^5$, for (a) random recursive trees (b) positive linear preferential attachment trees with $A = 0$ and (c) negative linear preferential attachment trees with $B = 3$. Lines represent fitted normal distributions with equal mean and standard deviation.}
\label{fig:P0_dist}
\end{figure}

The distribution of $P_0$, for all growth models and all parameters considered, appears to be independent of tree size (for large enough trees), as shown in Fig. \ref{fig:P0_dist}. For random recursive trees the distribution is well approximated by a normal distribution. For other growth models the distribution is distorted, but remains independent of size.

\begin{figure}[H]
\centering
\includegraphics[width=\columnwidth,angle=0.]{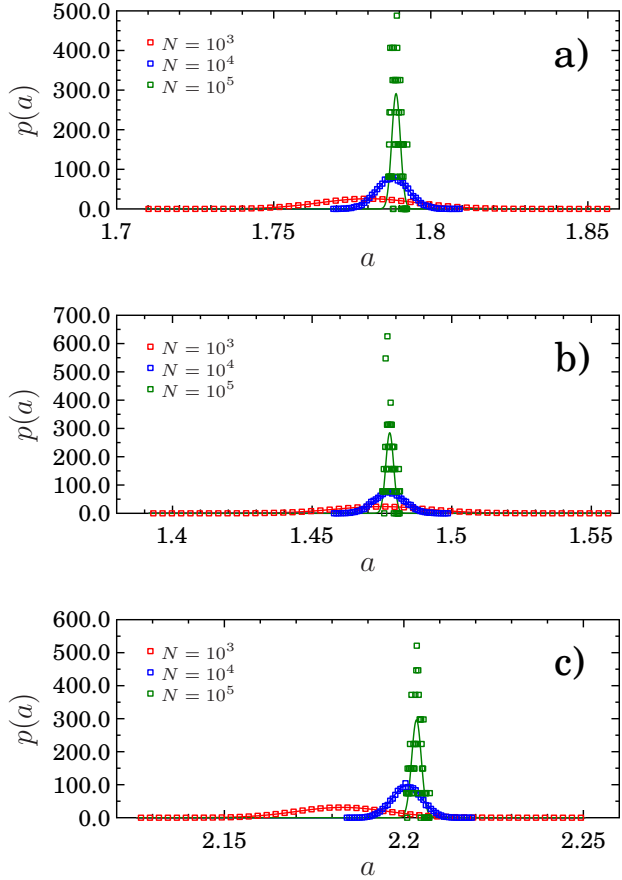}
\caption{(Color online). Distribution of $a$ values for three sizes, $N = 10^3$, $10^4$ and $10^5$, for (a) random recursive trees (b) positive linear preferential attachment trees with $A = 0$ and (c) negative linear preferential attachment trees with $B = 3$. Lines represent fitted normal distributions with equal mean and standard deviation.}
\label{fig:a_dist}
\end{figure}

\begin{figure}[H]
\centering
\includegraphics[width=\columnwidth,angle=0.]{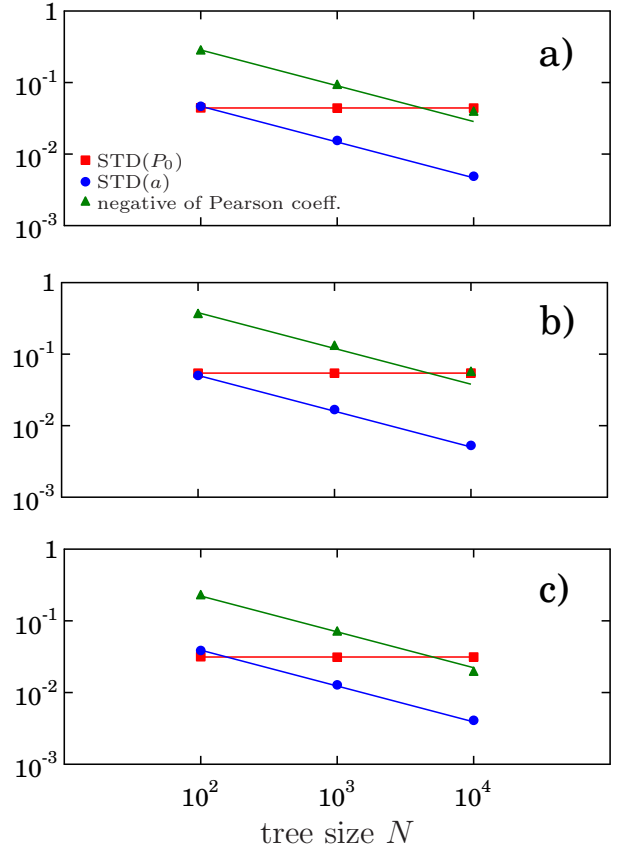}
\caption{(Color online). Standard deviation of $P_0$, $a$, and the negative of their Pearson correlation coefficient, as functions of system size. Red line is a constant, blue and green lines represent $\propto N^{-1/2}$. (a) shows random recursive trees, (b) positive linear preferential attachment trees with $A = 0$ and (c) negative linear preferential attachment trees with $B = 3$. Results were averaged over $10000$ realizations.}
\label{fig:correlations}
\end{figure}

The distribution of $a$ on the other hand appears to converge to a Dirac delta function with increasing size, see Fig. \ref{fig:a_dist}, and seems to be well-fitted by a normal distribution without the kind of distortions experienced by the distribution of $P_0$. Fig. \ref{fig:correlations} confirms that the standard deviation of $P_0$ is independent of system size, and the standard deviation of $a$ decays with system size approximately as $STD(a) \propto N^{-1/2}$.

For a given growth model and parameter setting, $a$ and $P_0$ appear to be negatively correlated for any size $N$, but the Pearson correlation coefficient appears to decay with size approximately as $\propto N^{-1/2}$ (see Fig. \ref{fig:correlations}). The fact that $a$ and $P_0$ are uncorrelated for large $N$ is particularly interesting because it implies that Eq. (\ref{eq:72.6}) does not derive from any simple relationship between two quantities related with a given tree, but rather, that it is a ``hidden'' property of the growth model class.

\input{paper.bbl}

\end{document}

%% file: paper.bbl
%